\newcommand{\beq}{\begin{equation}}
\newcommand{\eeq}{\end{equation}}
\newcommand{\bea}{\begin{eqnarray}}
\newcommand{\eea}{\end{eqnarray}}

\newcommand{\gsim}{\lower.7ex\hbox{$\;\stackrel{\textstyle>}{\sim}\;$}}
\newcommand{\lsim}{\lower.7ex\hbox{$\;\stackrel{\textstyle<}{\sim}\;$}}

\newcommand{\mrm}{\mathrm}

\documentclass[aps,prev,twocolumn,preprintnumbers,floatfix,nofootinbib]{revtex4-1}
\usepackage{graphicx}
\usepackage{epstopdf}
\usepackage{mathrsfs}
\usepackage{amssymb}
\usepackage{verbatim}
\usepackage{color}

\def\stacksymbols #1#2#3#4{\def\theguybelow{#2}
    \def\vp{\lower#3pt}
    \def\sp{\baselineskip0pt\lineskip#4pt}
    \mathrel{\mathpalette\intermediary#1}}

\def\intermediary#1#2{\vp\vbox{\sp
     \everycr={}\tabskip0pt
     \halign{$\mathsurround0pt#1\hfil##\hfil$\crcr#2\crcr
              \theguybelow\crcr}}}

\def\be{\begin{equation}}
\def\ee{\end{equation}}
\def\bea{\begin{eqnarray}}
\def\eea{\end{eqnarray}}

\def\sp{\;\;\;,\;\;\;}

\def\mrm{\mathrm}

\def\lsim{\raise0.3ex\hbox{$\;<$\kern-0.75em\raise-1.1ex\hbox{$\sim\;$}}}
\def\gsim{\raise0.3ex\hbox{$\;>$\kern-0.75em\raise-1.1ex\hbox{$\sim\;$}}}

\def\inbar{\,\vrule height1.5ex width.4pt depth0pt}

\def\IC{\relax\hbox{$\inbar\kern-.3em{\rm C}$}}
\def\IQ{\relax\hbox{$\inbar\kern-.3em{\rm Q}$}}
\def\IR{\relax{\rm I\kern-.18em R}}
 \font\cmss=cmss10 \font\cmsss=cmss10 at 7pt
\def\IZ{\relax\ifmmode\mathchoice
 {\hbox{\cmss Z\kern-.4em Z}}{\hbox{\cmss Z\kern-.4em Z}}
 {\lower.9pt\hbox{\cmsss Z\kern-.4em Z}}
 {\lower1.2pt\hbox{\cmsss Z\kern-.4em Z}}\else{\cmss Z\kern-.4em Z}\fi}

\def\comment#1{}

\def\u1x{U(1)_X}
\newcommand{\nc}{\newcommand}
\nc{\LL}{L}
\nc{\vv}{\tilde{v}}
\nc{\ccdot}{\!\cdot\!}
\nc{\gsm}{G_{SM}}
\nc{\vfive}{\mathbf{5}\oplus\mathbf{\overline{5}}}
\nc{\vten}{\mathbf{10}\oplus\mathbf{\overline{10}}}
\nc{\zhol}{Z^{\rm hol}}
\nc{\xfb}{\,{\rm fb}}

\begin{document}

%
%

\preprint{LPT--Orsay 11/47}
\preprint{IFT-UAM/CSIC-11-44}

\title{When LEP and Tevatron  combined with WMAP and XENON100 shed light on the nature of Dark Matter}
\author{Yann Mambrini$^{a}$}
\email{Yann.Mambrini@th.u-psud.fr}
\author{Bryan Zald\'\i var$^{b}$}
\email{Bryan.Zaldivar@uam.es}

\vspace{0.2cm}
\affiliation{
${}^a$ Laboratoire de Physique Th\'eorique 
Universit\'e Paris-Sud, F-91405 Orsay, France \\
${}^b$Instituto de Fisica Teorica, IFT-UAM/CSIC
Nicolas Cabrera 15, UAM
Cantoblanco, 28049 Madrid, Spain}

\begin{abstract}

Recently, several  astrophysical data or would-be signals  has been observed in different dark-matter
oriented experiments.
In each case, one could fit the data at the price of specific nature of the coupling between the Standard Model 
(SM) particles and a light Dark Matter candidate: hadrophobic
(INTEGRAL, PAMELA) or leptophobic (WMAP Haze, dijet anomalies of CDF, FERMI Galactic Center observation).
In this work, we show that when one takes into account the more recent LEP and Tevatron analysis, a light thermal 
fermionic Dark Matter ($\lesssim 10$ GeV) 
that couples to electrons is mainly ruled out
if one combines the analysis  with WMAP constraints. 
We also study the special case of scalar dark matter, using a mono--photon events simulation
to constrain the coupling of dark matter to electron.

\end{abstract}

\maketitle


\maketitle


\setcounter{equation}{0}



\section{Introduction}

Very recently, the CDF collaboration announced the observation of an excess of events which
 include a lepton (electron or muon), missing transverse energy, and
two jets \cite{Aaltonen:2011mk}. Many studies have been done since then motivating the existence of light dark matter candidates (see e.g., \cite{Hooper:2010uy}). Some authors \cite{Buckley:2011vs} interpreted 
 this excess by
the introduction of a new gauge boson with sizable couplings to quarks, 
 but with no or highly suppressed couplings to leptons (a $leptophobic$ dark boson).
 Dark matter experiments had also given some hints for signals in direct or indirect
 detection modes.
 On one hand, some hadrophobic dark matter candidates were proposed in 
 \cite{Bernabei:2007gr,Fox:2008kb,Dedes:2009bk} to explain
 the DAMA \cite{DAMA} and CoGENT \cite{COGENT} signals even if contradicted by the authors of
  \cite{Kopp:2009et,Kopp:2010su}. On the other hand, some authors showed that
 a light dark matter could at the same time explain these direct detection signals
 and  the excess of emission observed by the Fermi Gamma Ray Space telescope \cite{Hooper:2010mq}
 and the CDF signal if it annihilate predominantly into hadronic states.
 There was also cosmic rays excess measured in PAMELA or INTEGRAL \cite{Boehm:2003bt} which
 needed hadrophobic dark matter. 
 In each case the nature of the couplings of the dark matter with the Standard Model particles
 is fundamental in any kind of discoveries. Recently, the authors of \cite{Fox:2011fx} used the mono--photon
 events at LEP to constraint the nature of the dark matter couplings, concluding that a
 dark matter with mass $\lesssim 10$ GeV  with charged-leptonic couplings generates a too low annihilation
 rate to avoid the over--closure of the Universe.
 In this work, we compute the rate of hadronic coupling needed to reconciliate the LEP 
 analysis with a thermal dark matter hypothesis and respect WMAP upper bound constraint.
 In  section II, we will review the models and type of couplings we have studied. We give our result
 in the case of contact operator for a fermionic candidate in section IIIA, and consider a scalar case in section IIIB. 
 For the later, we ran a simulation of events at DELPHI experiment \cite{DELPHI:2005} in order to constraint  
 the operator suppression scale, in the same fashion as is done in the literature for the fermionic DM.  
 We then implement the constraints from the mono-jet event of Tevatron and XENON100 
 in the analysis in section IV before concluding in section V.

\section{The models}

We begin with the case of a fermionic WIMP, and study the 4 types of interactions
consistent with the requirement of Lorentz invariance and strongly constrained by LEP
analysis. This enables us to describe the interaction between WIMPs and 
standard model fermions in terms of an effective field theory, in which we keep only the first term
in the expansion of the heavy propagator. However, contrarily to the description in \cite{Fox:2011fx}
which was concerned by the leptonic constraints, we generalize the analysis taking into account
1) the neutrino couplings and 2) the possibility of hadronic tree level couplings.
This implies the introduction of a second effective scale, $\Lambda_h$. Indeed there is no reason for
the effective hadronic breaking scale to be the same than the leptonic one $\Lambda_l$. 
 We will thus introduce hadronic and leptonic coupling constants $g_h$ and $g_l$, such as 

\be
\frac{1}{\Lambda_l} \equiv \frac{\sqrt{g_l}}{\Lambda}; ~~~~~ 
\frac{1}{\Lambda_h}  \equiv \frac{\sqrt{g_h}}{\Lambda}
\ee
 
We will then consider the set of operators

\bea
&&
\mrm{Vector : }~  {\cal L}_V = \sum_i \frac{g^i_l}{\Lambda^2}( \bar l^i \gamma^\mu l^i )(\bar \chi \gamma_\mu \chi)
+ \sum_i \frac{g_h^i}{\Lambda^2}( \bar q^i \gamma^\mu q^i)( \bar \chi \gamma_\mu \chi)
\nonumber
\\
&&
\mrm{Scalar, s-channel:}
 ~ {\cal L}_S = \sum_i \frac{g^i_l}{\Lambda^2} (\bar l^i  l^i )(\bar \chi  \chi)
+ \sum_i \frac{g_h^i}{\Lambda^2} (\bar q^i  q^i)( \bar \chi  \chi)
\nonumber
\\
&&
\mrm{Axial : }~  {\cal L}_A = \sum_i \frac{g^i_l}{\Lambda^2} (\bar l^i \gamma^\mu \gamma^5 l^i)( \bar \chi \gamma_\mu 
\gamma^5 \chi) \nonumber \\
&&
\hspace{3.5cm}+ \sum_i \frac{g_h^i}{\Lambda^2} (\bar q^i \gamma^\mu \gamma^5 q^i)( \bar \chi \gamma_\mu \gamma^5 \chi)
\nonumber
\\
&&
\mrm{Scalar, t-channel}
 ~ {\cal L}_t = \sum_i \frac{g^i_l}{\Lambda^2}( \bar l^i  \chi) ( \bar \chi  l^i)
 + \sum_i \frac{g_h^i}{\Lambda^2}( \bar q^i  \chi) (\bar \chi  q^i)
\nonumber
\\
\eea

 \noindent
 $\chi$ being the DM candidate. Throughout this paper, we will assume that the dark matter particle $\chi$
 is a Dirac fermion (except in section IIIB, where we consider a real scalar DM candidate).
 A vectorial interaction is motivated by the exchange of a $Z_{\mu}'$ \cite{Feldman:2007wj}
  whereas scalar interaction  is motivated by Higgs-portal like models \cite{Andreas:2010dz}.
 
 We will consider 3 kinds of models which could be representative  of UV completion:
 
 \begin{itemize}
 \item{Electrophilic couplings (model A): $g_l^e=g_e$, $g_l^{i=\mu,\tau,\nu_i}=0$}
 \item{Charged lepton couplings (model B) : $g_l^{i=e,\mu,\tau}=g_l $, $g_l^{i=\nu_i}=0$}
  \item{Universal lepton couplings (model C) : $g_l^{i=e,\mu,\tau,\nu_i}=g_l$}
 \end{itemize}
 
 \noindent

 As we are interested in the ratio of the hadronic to the leptonic final states in the DM annihilation,
 we will consider without loss of generality an universal generation/family coupling in the hadronic sector:
 $g_h^{i=u,d,c,s,b,t}=g_h$. Note that we assumed lepton flavor to be conserved in the dark matter 
 interaction. 
 
\begin{figure}
    \begin{center}
   \includegraphics[width=3.in]{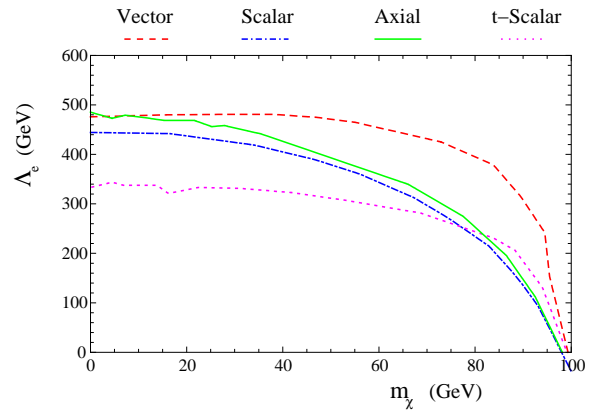}
   
          \caption{{\footnotesize
DELPHI lower limit on $\Lambda_e \equiv \Lambda/\sqrt{g_e}$ as a function of the dark matter mass for the
different types of couplings : vector (red dashed), scalar (blue dashed-dotted), axial (green full--line) and
t-channel scalar (magenta dotted).
}}
\label{Fig:LEP}
\end{center}
\end{figure}

 Recently, the authors of \cite{Fox:2011fx} made an analysis with relatively little model dependance,
  by pair production of pair of dark matter particles in association with a hard photon. The LEP experiments
have searched for anomalous mono--photon events in their data sets, but have found no discrepancy 
from the prediction of the standard model. They used the mono--photon spectrum from the DELPHI
experiment to place upper bound to $1/\Lambda_e^2=g_e/\Lambda^2$. We reproduce  interpolated functions 
of this result in Fig.(\ref{Fig:LEP}). We then translated this limit on $\Lambda_e$ to a limit 
on the ratio of hadronic to leptonic channel, $\mrm{Br_h/Br_l}$, taking into account the relic density constraints.

\section{Constraints from the thermal relic to the hadronic branching ratio}
\subsection{The fermionic case}

The LEP lower bound on the scale $\Lambda_e=\Lambda/\sqrt{g_e}$ can be converted in a upper 
bound to the dark matter annihilation into $e^+ e^-$ (case A), into charged leptons pair (case B)
or into general leptons pair (case C). Moreover, if dark matter is a thermal relic, asking for the 
density to respect the upper bound given by WMAP \cite{WMAP} $\Omega_{\chi} h^2 \lesssim 0.1$, one needs to
impose\footnote{in the absence of resonances or coannihilation.}
 $\langle \sigma v \rangle \gtrsim  3\times 10^{-26} \mrm{cm^3 s^{-1}} \simeq 1~ \mrm{pb}$ to avoid an overclosed universe 
(but letting for the possibility of having another dark matter candidate). We computed the annihilation
cross section which is given for a final state with particles masses $m_3, m_4$ by

\be
\frac{d\sigma_\mrm{I}}{d\Omega}=\frac{|{\cal M_\mrm{I}}|^2}{64 \pi^2 s}
\frac{\sqrt{s - 2m_3^2 -2m^2_4+\frac{(m^2_3-m^2_4)^2}{s}}}{\sqrt{s-4 m^2_{\chi}}}
\label{Eq:Sigma}
\ee

\noindent
with I=V, S, A, t.
We then substitute $s \simeq 4 m^4_{\chi} + m^2_{\chi} v^2$ in Eq.(\ref{Eq:Sigma}) 
and expanding in powers of the relative velocity
between two annihilating WIMPs up to order $v^2$ for each type of couplings. We find

\bea
\sigma^J_I v = g_l^2 \sum_{l=e,\mu,\tau,\nu}\sigma^J_{I,l} v 
+c~ g_{h}^2 \sum_{h=u,d,c,s,t,b}\sigma^J_{I,h} v 
\eea

\noindent 
where I=V, S, A, t represents the nature of the coupling (vectorial, scalar, axial or t-scalar)
 and J=A, B, C the type of coupling (electronic, 
charged leptonic or universal leptonic), $c$ the color factor and

\bea
&&
\sigma^J_{Ik} v = \sigma_{I,k} v \times \theta^J_k(m_{\chi}) 
\\
&&
\theta^{J=A,B,C}_{h=u,d,c,s,t,b}(m_{\chi}) = \Theta_H (m_{\chi}-m_h)
\nonumber
\\
&&
\theta^A_e (m_{\chi})= \Theta_H (m_{\chi}-m_e), ~~~ \theta^A_{l=\mu, \tau,\nu}(m_{\chi}) = 0
\nonumber
\\
&&
\theta^B_{l=e,\mu,\tau} (m_{\chi})= \Theta_H (m_{\chi}-m_l), ~~~ \theta^B_{l=\nu}(m_{m_{\chi}}) = 0
\nonumber
\\
&&
\theta^C_{l=e,\mu,\tau,\nu}  (m_{\chi})= \Theta_H (m_{\chi}-m_l)
\nonumber
\eea

\noindent
$\Theta_H$ being the classical heaviside function ($\Theta_H(x)=1$ if $x>0$, and 0 otherwise) and
$\sigma_{I,k}$ is given by:

\bea
&&
\sigma_{V,k} v =4 g_{\Lambda}
\left(
24(2 m_{\chi}^2+m_k^2)
+\frac{8 m_{\chi}^4 - 4 m_{\chi}^2 m_k^2 + 5 m_k^4}{m^2_{\chi}-m_l^2}v^2
\right)
\nonumber
\\
&&
\sigma_{S,k}v =24 g_{\Lambda}
(m_{\chi}^2 - m_k^2)v^2
\label{Eq:Sigmav}
\\
\nonumber
&&
\sigma_{A,k} v =4 g_{\Lambda}
\left(
24 m_k^2 +\frac{8 m_{\chi}^4 - 22 m_{\chi}^2 m_k^2 + 17 m_k^4}{m_{\chi}^2-m_k^2}
v^2
\right)
\nonumber
\\
&&
\sigma_{t,k} v = g_{\Lambda}
\left(
24(m_{\chi} + m_k)^2
\nonumber
\right.
\\
&&
+
\frac{(m_{\chi}+m_k)^2(8 m_{\chi}^2 -16 m_{\chi}m_k+11 m_k^2)}{m_{\chi}^2-m_k^2} v^2
)
\nonumber
\eea

\noindent
with $g_{\Lambda}=\frac{\sqrt{1-m_{k}^2/m_{\chi}^2}}{192 \pi \Lambda^4}$.

\begin{figure}
    \begin{center}
   \includegraphics[width=3.in]{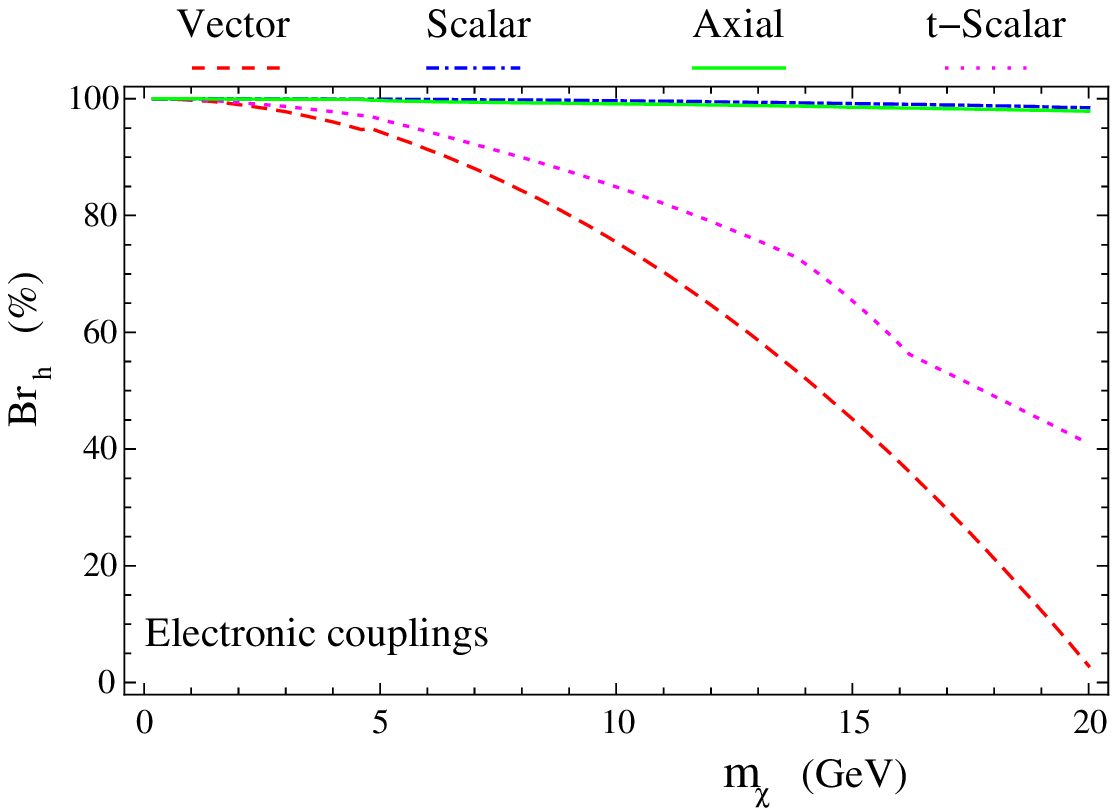}
   
    \includegraphics[width=3.in]{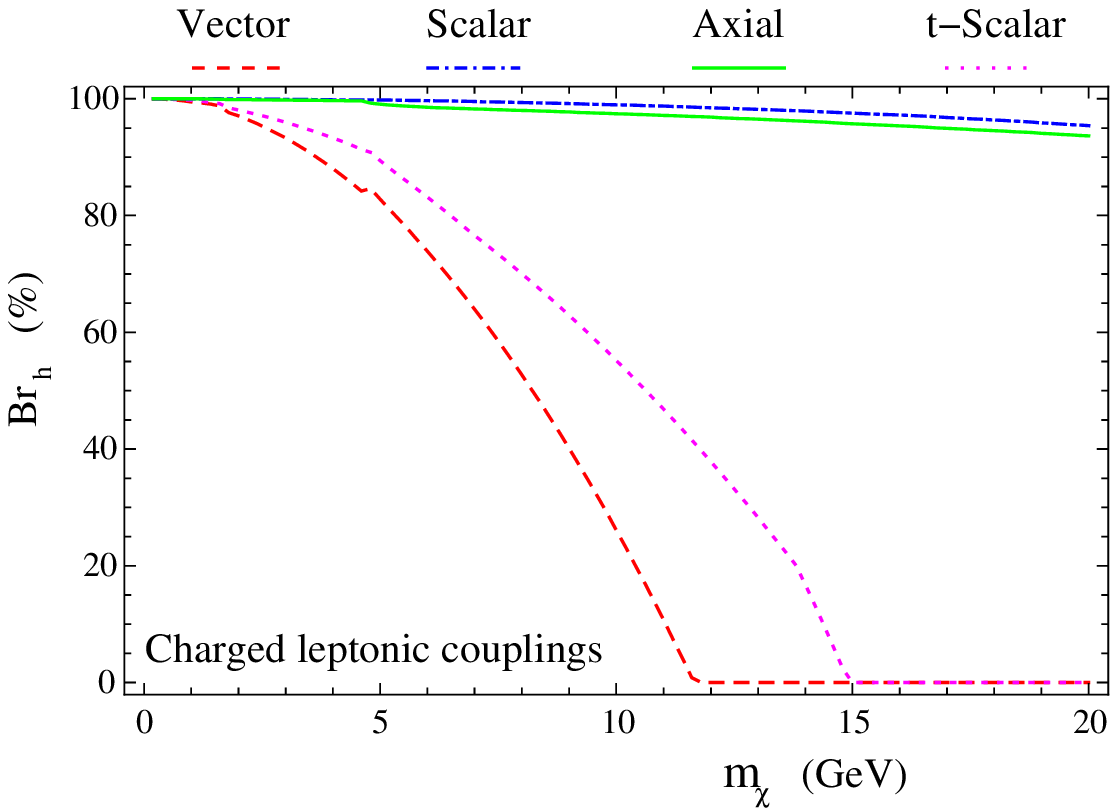}
    
    \includegraphics[width=3.in]{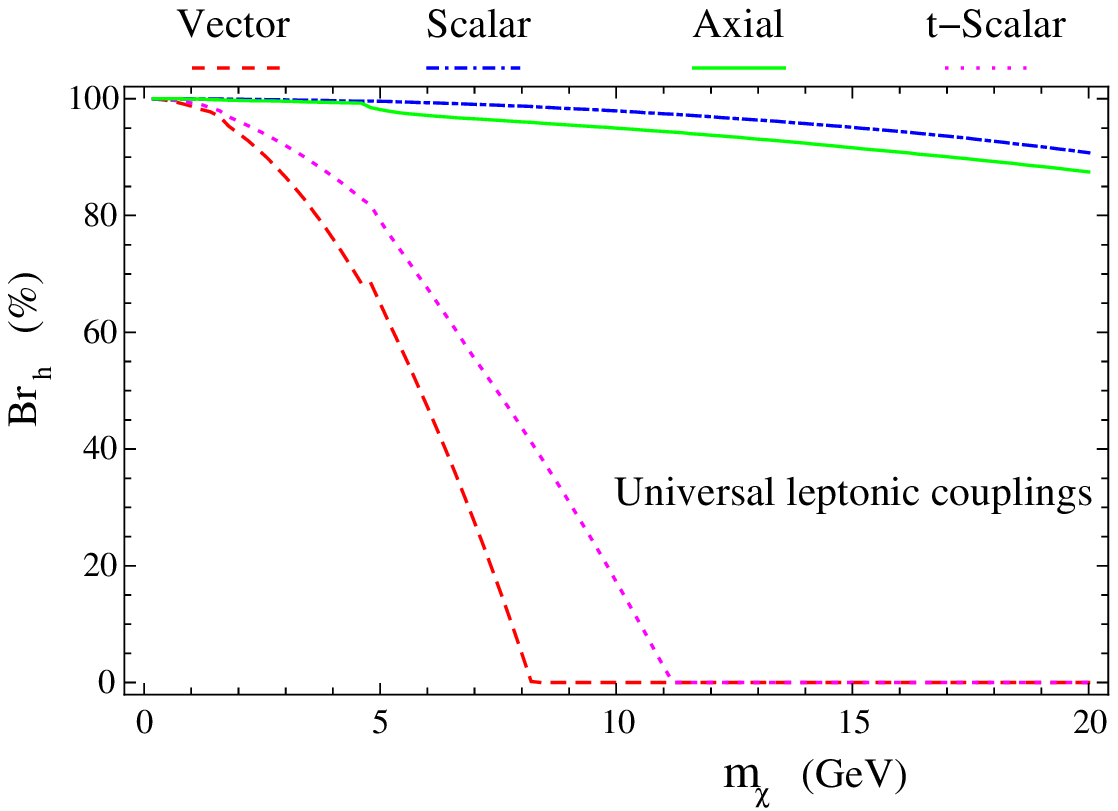}
          \caption{{\footnotesize
Minimum hadronic branching ratio needed to respect WMAP upper bound in the case of electronic couplings 
(model A, top), charged-leptonic couplings (model B, middle) and universal-leptonic couplings
(model C, bottom) with 4 different types of interactions: vector (red dashed), scalar (blue dashed-dotted),
 axial (green full--line) and
t-channel scalar (magenta dotted). Bounds coming from LEP constraints on leptonic couplings.
}}
\label{Fig:Contact}
\end{center}
\end{figure}

\noindent
The LEP constraint on $\Lambda_e=\Lambda/\sqrt{g_e}$ gives a maximum 
value for the leptonic annihilation cross section $\sigma^{max}_{l}$ for each type of couplings
we considered (A, B and C, see Fig.\ref{Fig:LEP}). This maximum value of the leptonic cross--section give a lower bound
on $\Omega_{\chi}h^2$ : one thus can calculate the hadronic contribution needed to satisfy
WMAP upper bound limit ($\Omega_{\chi}h^2 \lesssim 0.1$)
corresponding to the thermal condition $\sigma v \gtrsim 3 \times 10^{-26}\mrm{cm}^3$/s. This can be summarize 
by:

\be
\sigma^{max}_l v + \sigma^{max}_h v \gtrsim 3 \times 10^{-26} \mrm{cm^3s^{-1}}
  \simeq 2.5 \times 10^{-9} ~ \mrm{GeV^{-2}}
\label{Eq:WMAP}
\ee

As an example, we can analytically evaluate the order of magnitude for the hadronic branching ratio
$Br_h/Br_l$
we expect for a dark matter mass $m_{\chi} \simeq 5$ GeV
  in the case of an
electronic (case A) vector--like coupling (${\cal L}_V$). We combined the condition given in 
Eq.(\ref{Eq:WMAP}) to the value of $\sigma_V v$ computed through Eq.(\ref{Eq:Sigmav}) with
the value of $\Lambda_e$ obtained by LEP (see Fig.\ref{Fig:LEP}):
$\Lambda^{max}_{eV} \simeq 480$ GeV for $m_{\chi} \simeq 5$ GeV. Neglecting
$m_{l,h} \ll m_{\chi}$, in the electronic--type coupling, 
one can simplify 
$\sigma_V v \simeq \frac{m^2_{\chi}}{\pi \Lambda_e^4}\left( 1+ Br_h/Br_l \right)\gtrsim 2.5 \times 10^{-9}$
which gives
$Br_h/Br_l \gtrsim \frac{(\Lambda^{max}_e)^4 \pi}{m^2_{\chi}} 2.5 \times 10^{-9} \simeq 16$.
This corresponds to a 94\% annihilation rate to hadronic states. Of course, we ran the analysis with the complete
formulation for the cross sections and the results are shown in Fig.\ref{Fig:Contact}.
One can see that whatever is the nature of the coupling (electronic, charged-leptonic or universal leptonic), a dark matter of mass
 $m_\chi\simeq$ 10 GeV has a very strong  hadronic component in its annihilation final state, in the case of scalar and axial interactions
 (above 90$\%$). On the other hand, for vector and t-scalar interactions, the nature of the coupling plays an important role, being 
an hadronic component as large as 80$\%$ in one case (electronic coupling), or a 0$\%$ (i.e. no need of hadronic channel) 
in other case (universal leptonic), for a vector interaction for example. 
These behaviors can be understood from expressions (\ref{Eq:Sigmav}), where the scalar and axial interactions are suppressed by the velocity and the leptonic masses, respectively. As a consequence one needs a much larger hadronic contribution to ensure
a relic abundance below the WMAP limit and avoid the over--closure of the Universe. However for the vector and t-scalar interactions there is no such suppression, so leading to possible large contributions coming only from the leptonic couplings.
  
We also observe that, paradoxically,   the more electrophilic are the dark matter couplings (model A), the more hadrophilic
it should also be. Indeed, because there are no possibility to fulfill the relic abundance constraints with charged lepton
or neutrino channels and the hadronic final states become thus the dominant ones. 
In the charged leptonic and universal leptonic models (B and C), there exists a threshold mass
with a null hadronic branching ratio: this corresponds to the mass for which the hadronic components of
the annihilation rate are not anymore necessary (but can be present) to fulfill the relic density constraints. The
leptonic channels are sufficient to avoid the relic overabundance for a DM mass above this threshold.

\subsection{Scalar case}

\begin{figure}
    \begin{center}
       \includegraphics[width=3.5in]{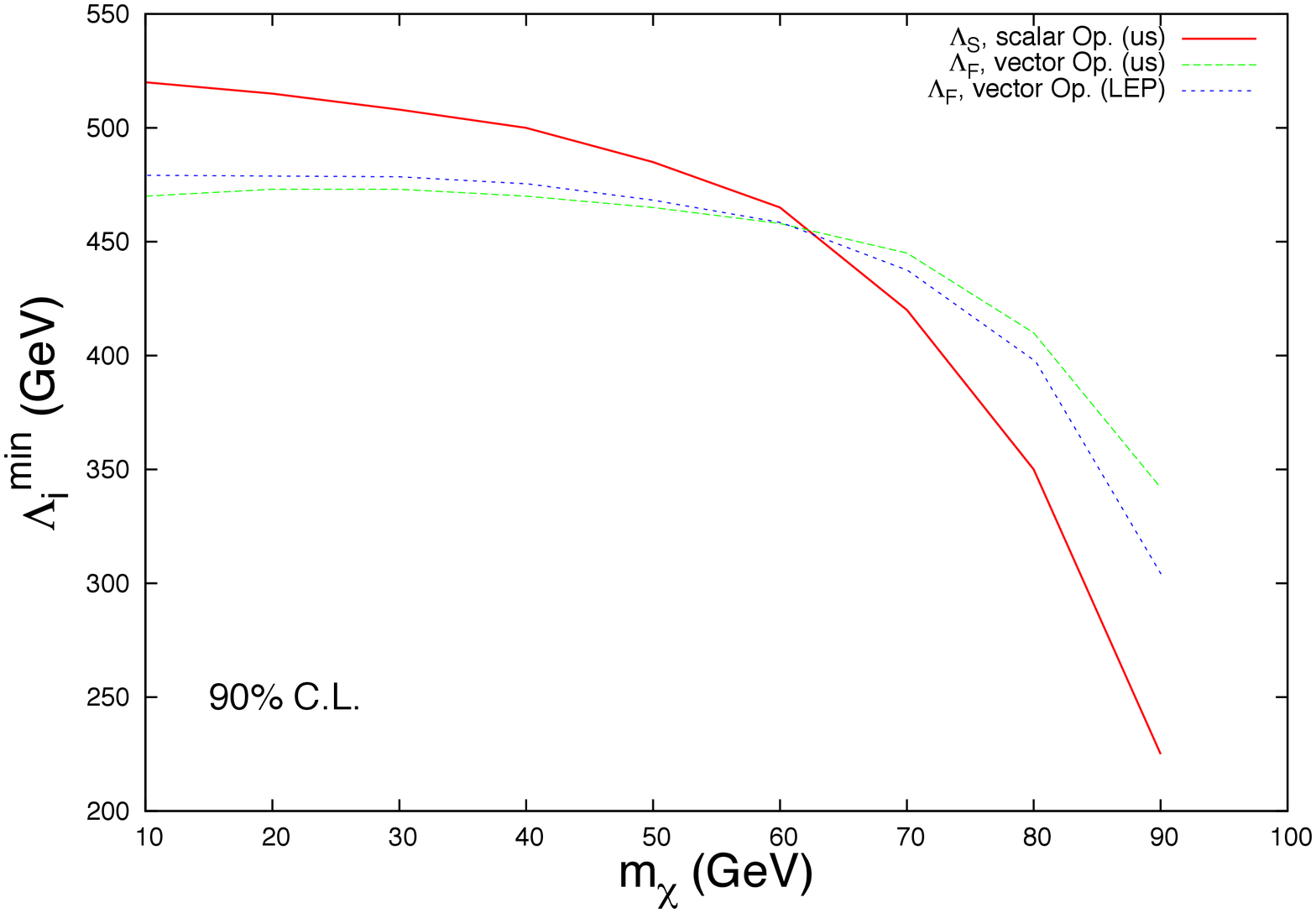}
   \includegraphics[width=3.5in]{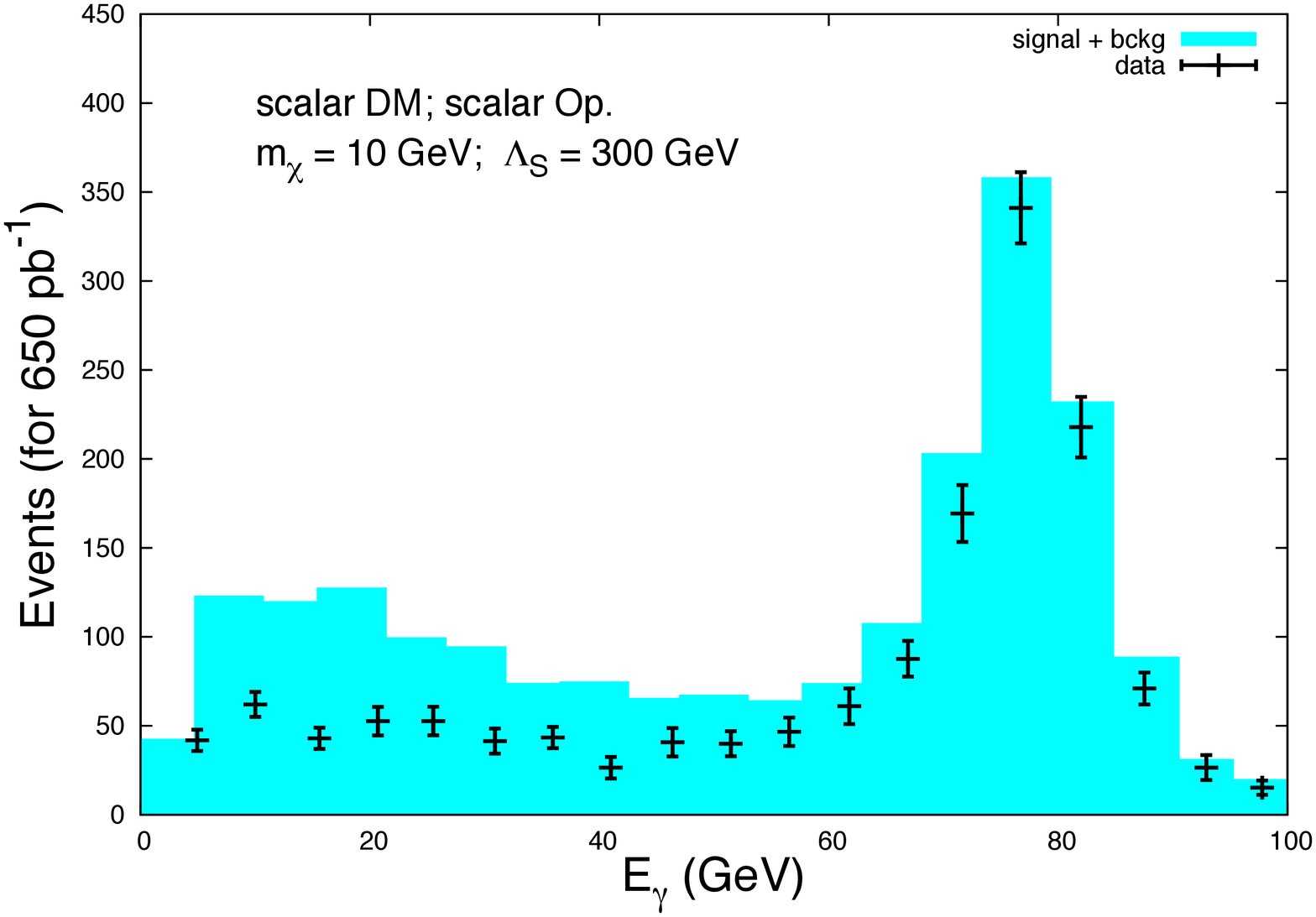}
         \caption{{\footnotesize
(top).  Lower limits on $\Lambda_S/g_e$ (solid-red) coming from DELPHI experiment \cite{DELPHI:2005}, at a 90$\%$ C.L. For reference, the resulting limits on $\Lambda/\sqrt{g_e}$ (fermionic dark matter) coming from a vector-like effective operator, using the same cuts as before, are shown (dashed-green), to be compared with the correspondent result shown in Fig.\ref{Fig:LEP}, here in (dotted-blue). (bottom) Distribution of photon energies in single-photon events at DELPHI. The histogram shows the signal$+$background coming from a hypothetical scalar dark matter, as in (\ref{scalarDM}), with mass $m_\chi = 10$ GeV, and a suppression scale $\Lambda_S/g_e = 300$ GeV.  See body text.
}}
\label{Fig:Scalar}
\end{center}
\end{figure}

We also checked the case of a scalar dark matter.
It could not be obvious at the first sight that we can apply the same analysis. In fact, we need to
introduce a new scale $\Lambda_S$. We will consider a real scalar dark matter, which is produced via the following scalar-type effective operator: 
\be
{\cal L}^S_e= \frac{g_e}{\Lambda_S}\chi \chi \bar e e ~.
\label{scalarDM}
\ee
In an analogous procedure as the one done above for a fermionic candidate, we derive limits on the suppression scale $\Lambda_S/g_e$ from 
mono-photon signals in the DELPHI experiment at LEP. We assumed for simplicity that all data was taken at an energy of 100 GeV
 per beam. We used MadGraph/MadEvent \cite{madgraph} to simulate the distribution of number of events with photon energy 
 $E_\gamma$. The background process $e^+ e^-\rightarrow \gamma \nu\bar\nu$ was taken directly from the simulation done 
 by \cite{Fox:2011fx}. On the other hand, the signal process $e^+ e^-\rightarrow \gamma\chi\chi$ was studied assuming the
  following kinematical cuts: $E_\gamma > 6$ GeV, and a photon rapidity $\eta_\gamma > 2.5$. We realize that these constraints 
  are less restrictive than in the fermionic case,
   so in principle the bounds on $\Lambda_S$ could be different if using those more rigorous cuts. 
To quantify this difference, we reproduced the  bounds on $\Lambda_e$  coming from a signal due
 to a fermionic dark matter, and a vector-like effective operator, and compare it directly with the result shown in Fig.\ref{Fig:LEP}. 
 The result, shown in Fig.\ref{Fig:Scalar}-top, is a $\chi^2$/d.o.f. $= 5.12/8$, which means a small difference of our case with respect 
 to the more correct result of \cite{Fox:2011fx}.  We include as an example of a scalar dark matter signal, the simulation of a 
 $m_\chi = 10$ GeV case, with a suppression scale $\Lambda_S/g_e = 300$ GeV, using the DELPHI luminosity of 650 pb$^{-1}$, 
 shown in Fig.\ref{Fig:Scalar}-bottom.  We can extract from the above analysis that, for example, a scalar dark matter 
 of $m_X=10$ GeV needs a suppression scale $\Lambda_S \gtrsim 520$ GeV, in order to be compatible with LEP bounds. 
 
\noindent
After computing the production cross section for the fermionic dark matter ($\sigma^P_f$) and bosonic one
($\sigma^P_s$) one can show that, in the limit of $s \gg m_{\chi}$ (which is the case in our analysis)
we obtained

\be
\sigma_f^P \sim \frac{g_e^2}{\Lambda^4}\frac{s}{16 \pi} ~~~~~~
\mrm{and}~~~~~
\sigma^P_s \sim \frac{g_e^2}{\Lambda_S^2}\frac{1}{32 \pi}
\ee

\noindent
We thus observe that if we define $\Lambda_S\equiv \Lambda^2/\sqrt{2 s}$ we can deduce the 
lower limit on $\Lambda_S$ from the lower limit on $\Lambda$ 
(see Fig.\ref{Fig:LEP}). If imposing $\sigma_f^P \approx \sigma_s^P$, taking a dark matter candidate with mass of 10 GeV, 
we deduce from the above expressions the lower bound $\Lambda_S\gtrsim 815$ GeV. However from Fig.(\ref{Fig:Scalar}) 
we obtained $\Lambda_S\gtrsim 520$ GeV, which implies that in fact $\sigma^P_s \approx 2.45 \sigma^P_f$. 

With this bounds on $\Lambda_S$, one could in principle try to deduce bounds on the amount of hadronic channel from DM annihilation, as we did above for the fermionic case. The expression for the annihilation cross-section $\sigma^s_S v$ of a scalar DM with a scalar interaction, into an electron-positron pair is:
\be
\sigma^s_{S,e} v  \equiv g_e^2 \tilde\sigma^s_{S,e} v \simeq \frac{g_e^2}{4\pi \Lambda_S^2} \left( 1- \frac{m_e^2}{m_\chi^2}\right)^{3/2} + \frac{g_e^2}{32\pi \Lambda_S^2} v^2 ~.
\ee 
Unfortunately, this single channel already gives, for $510 \lesssim \Lambda_S / g_e \lesssim 520$GeV  
and $1\lesssim m_\chi \lesssim 20$GeV (as in Fig.\ref{Fig:Scalar}) a cross-section $\sigma^s_S v  \simeq 10^{-24}\mrm{cm}^3/$s, 
with negligible dependence on $m_\chi$. Being $\sigma^s_S v \gg 3\times10^{-26}$, there is in principle no need for hadronic
 channel. We conclude that LEP bounds are insufficient to constrain the nature of couplings in the case of scalar DM.
  A similar conclusion holds for the Tevatron bounds, if considering the total cross-section 
\be
\sigma^s_{S} v = g_l^2 \sum_{l=e,\mu,\tau,\nu}  \tilde\sigma^s_{S,l} v + c g_h^2 \sum_{h=u,d,c,s,t,b}  \tilde\sigma^s_{S,h} v
\ee
and the lower limits shown in Fig.\ref{Fig:TEVATRON}. It turns out that the scale $\Lambda_S/g_e$ above which $\sigma^s_{S} v$ starts to be of the order of the thermal relic one, is around 5 TeV. So in principle the LHC would be able to constrain the nature of couplings and interactions of a scalar DM candidate. 

\section{Complementarity with other experiments}

\subsection{Fitting with WMAP}

To run a more precise analysis, we decided to implement the contact coupling lagrangian 
into CompHEP and micromegas \cite{Micromegas} for the different type of interactions
(vectorial, scalar and axial) in the fermionic DM case. We then applied the last $5\sigma$ constraint on the
relic density from WMAP experiment \cite{WMAP},
 $\Omega_{\mrm{WMAP}} h^2 = 0.1123\pm 0.0175 $ and ran a scan on the parameter space of the model 
($\Lambda_l$, $\Lambda_h$, $m_{\chi}$) keeping only the points respecting both astrophysical
and accelerator constraints.
We can understand easily that in order to respect WMAP upper bounds, the hadronic contribution 
 depends on 
the type of interactions. After a look at Fig.\ref{Fig:Contact}, we decided to consider the most and the less conservative cases, which are the universal-leptonic coupling for vector-like interaction, and the electronic coupling for scalar-like interaction, respectively.  The results are shown in Figs.\ref{Fig:TevatronVector} and \ref{Fig:TevatronScalar}.
 We see that WMAP forbids a dark matter with hadrophilic couplings ($g_h/g_e \gtrsim 10$) of $m_\chi \gtrsim 5$ GeV for a vector interaction, and $m_\chi \gtrsim 10$ GeV for a scalar interaction. 
 For such values of hadronic couplings, $\Omega h^2 \lesssim \Omega_{\mrm{WMAP}} h^2$.
 When combined with LEP analysis, a large part of the parameter space with small
 $g_h/g_e$ is excluded because of the non--observation of mono--jet events at LEP
 (which implies an upper bound on $g_e$). Whereas it excludes a broad region of
 the parameter space for a dark matter mass
 $m_{\chi}\lesssim 11$ GeV in the vectorial case (in total agreement with Fig.6 of \cite{Fox:2011fx})
 it completely exclude leptophilic ($g_h/g_e \lesssim 0.1$) dark matter with the scalar-like interaction.
 Combining these limits with the recent Tevatron analysis restricted even further the parameter space.

\subsection{Tevatron constraints}

\begin{figure}
    \begin{center}
   \includegraphics[width=3.in]{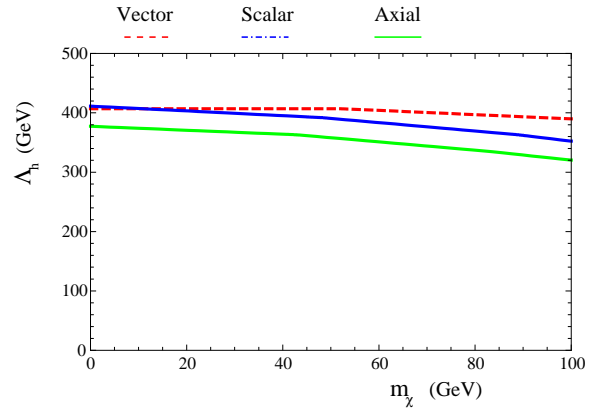}
   
          \caption{{\footnotesize
CDF lower limit on $\Lambda_h \equiv \Lambda/\sqrt{g_h}$ as a function of the dark matter mass for the
different types of couplings : vector (red dashed), scalar (blue dashed-dotted) and axial (green full--line).
}}
\label{Fig:TEVATRON}
\end{center}
\end{figure}

Last year, the authors of \cite{Bai:2010hh,Goodman:2010ku}
 made a similar analysis searching for mono--jet events for the Tevatron.
 These non-discovery of any events of this kind can be translated into a lower bound on 
$\Lambda_h \equiv \Lambda/\sqrt{g_h}$ which depend on the nature of the coupling and is represented
in Fig.\ref{Fig:TEVATRON}.

\begin{figure}
    \begin{center}
   \includegraphics[width=3.in]{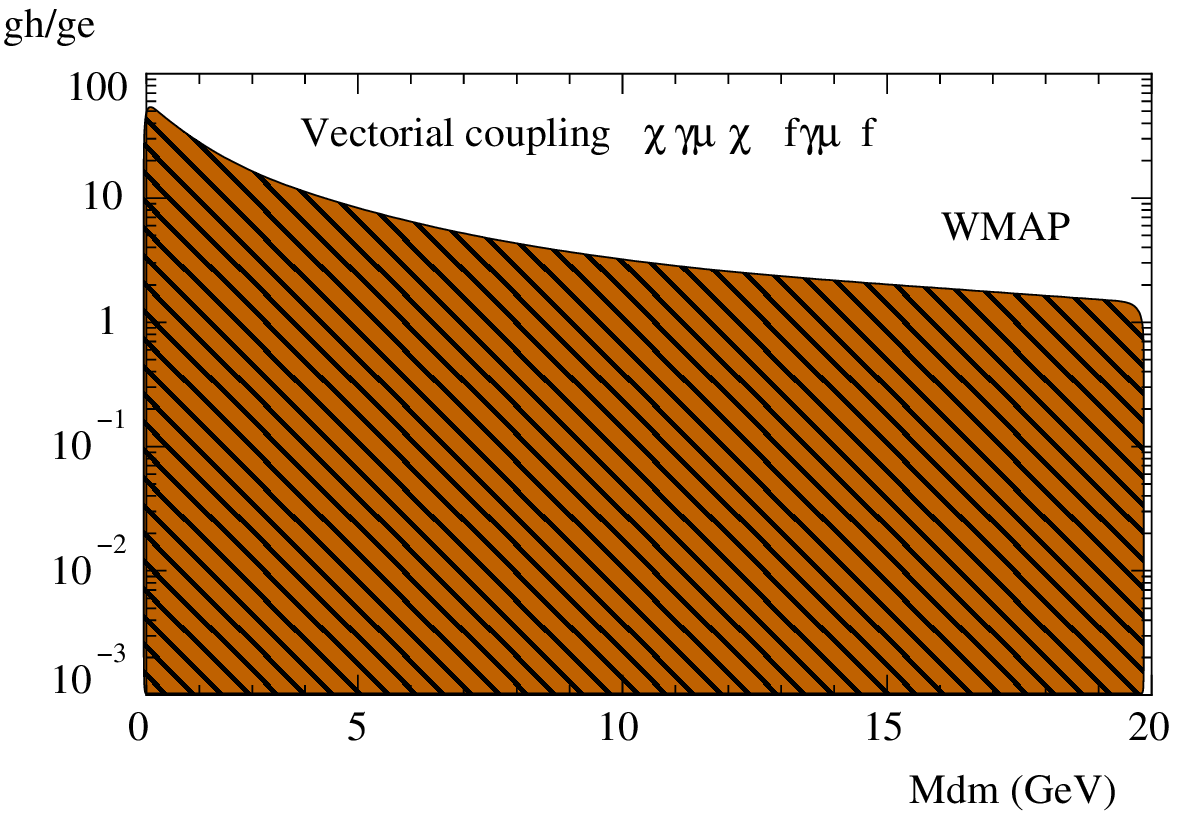}
   
    \includegraphics[width=3.in]{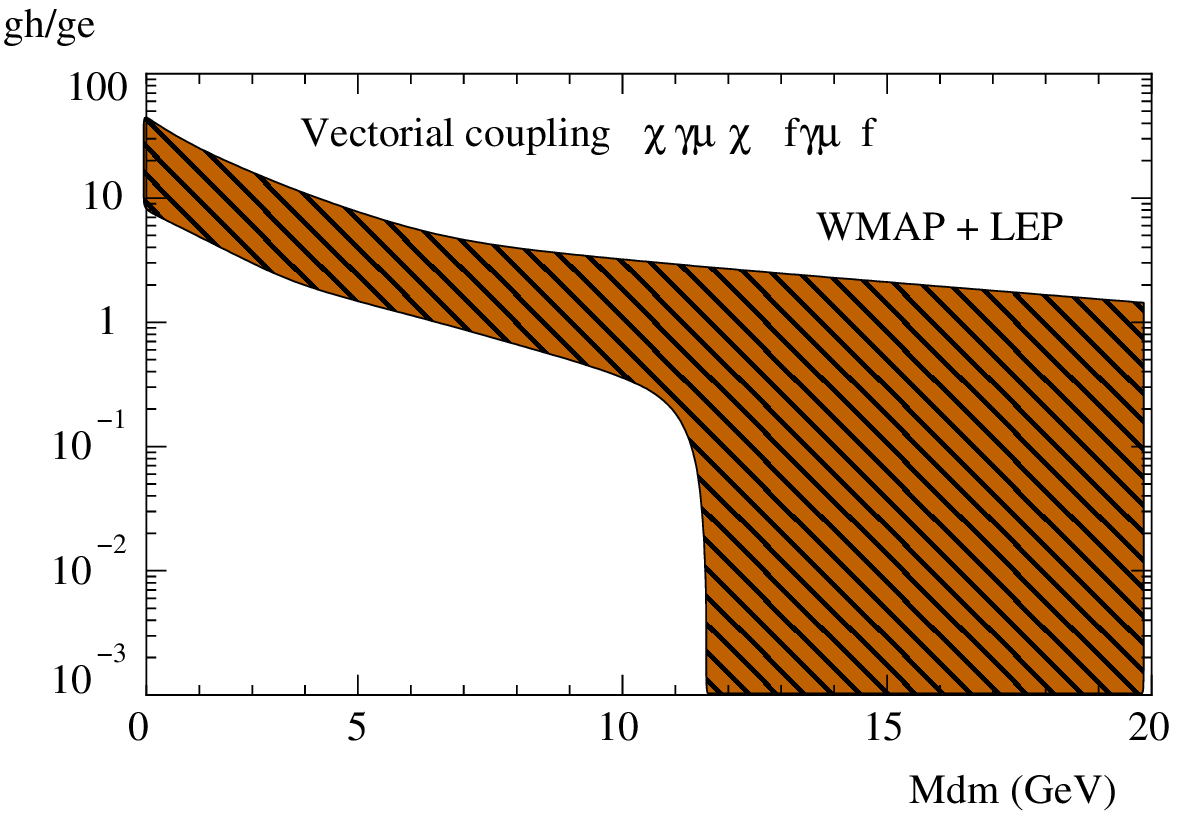}
    
       \includegraphics[width=3.in]{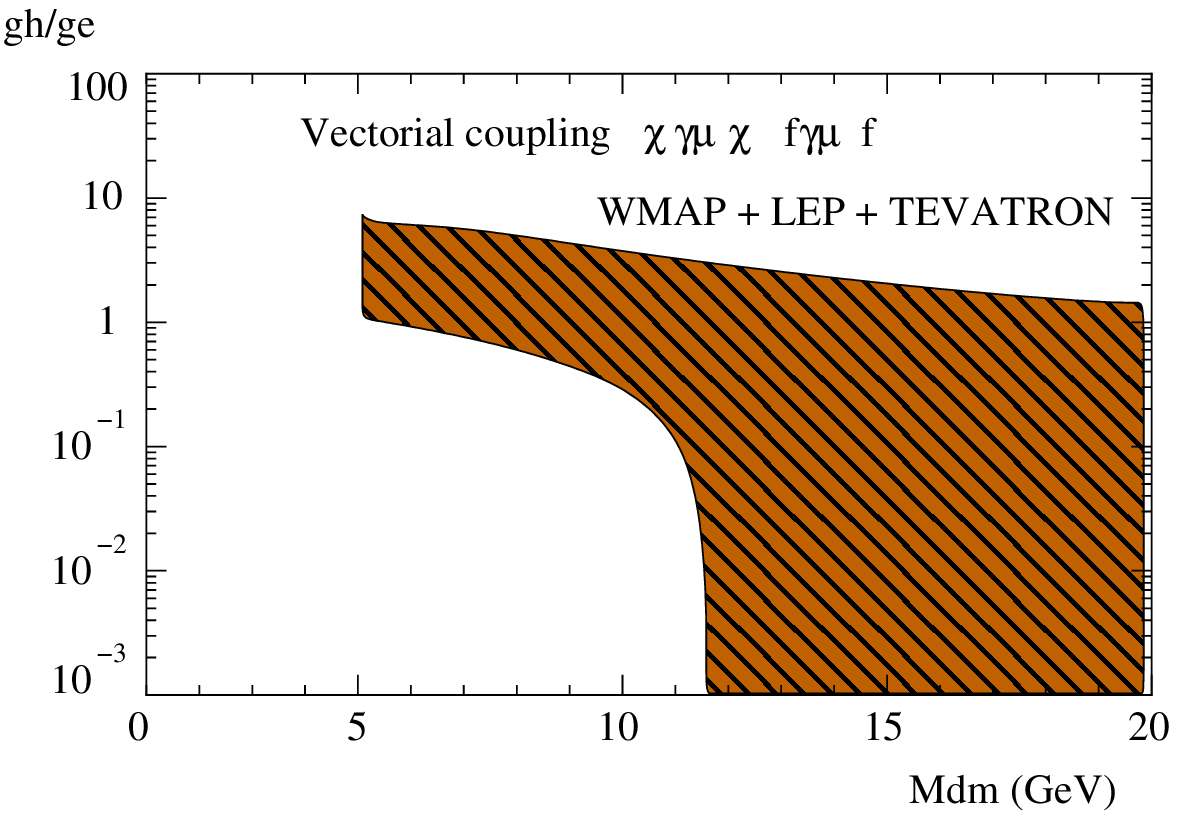}
   
    \includegraphics[width=3.in]{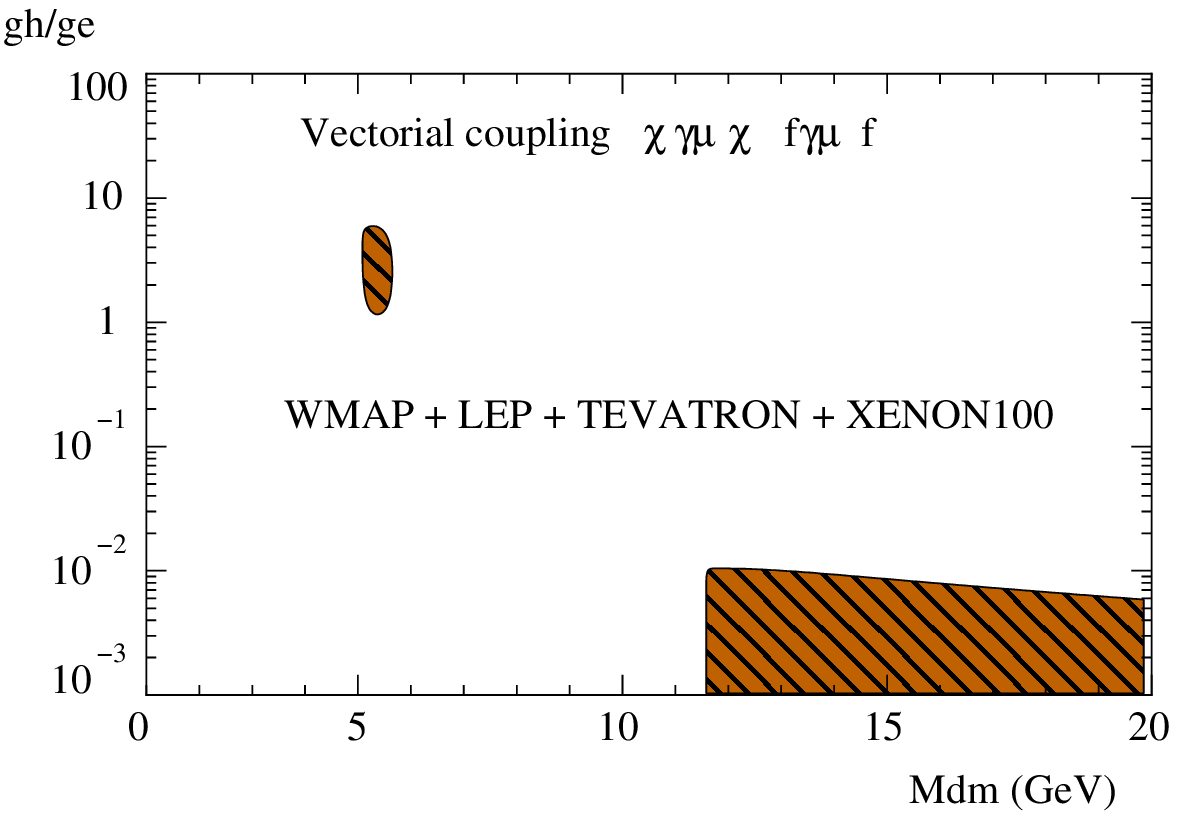}

          \caption{{\footnotesize
Hadronic ratio coupling for the annihilation of dark matter as function of the dark matter mass
in the case of universal-leptonic couplings for a vector--like interaction after a scan on $\Lambda_e$
and $\Lambda_h$. After applying the constraint of WMAP (top) mono--jet events from LEP, Tevatron
and XENON100 constraint (bottom). See the text for details.
}}
\label{Fig:TevatronVector}
\end{center}
\end{figure}

\begin{figure}
    \begin{center}
   \includegraphics[width=3.in]{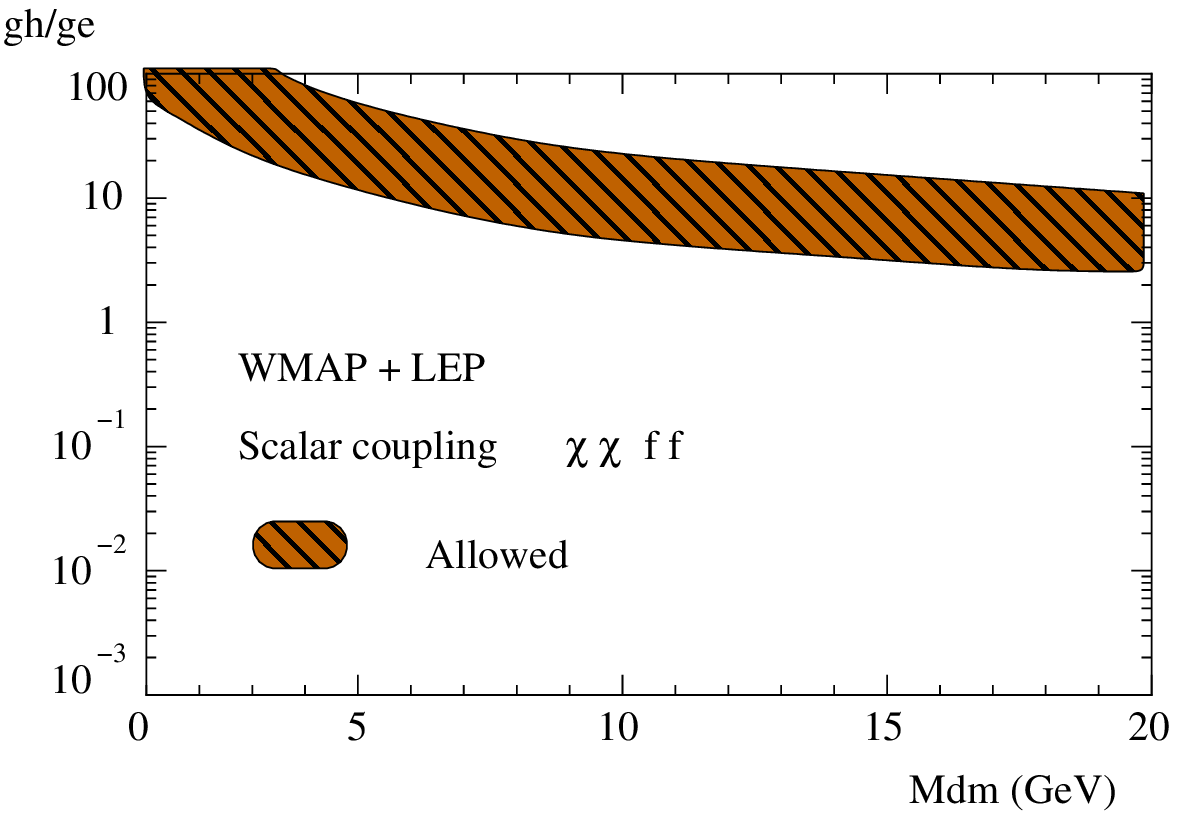}
   
    \includegraphics[width=3.in]{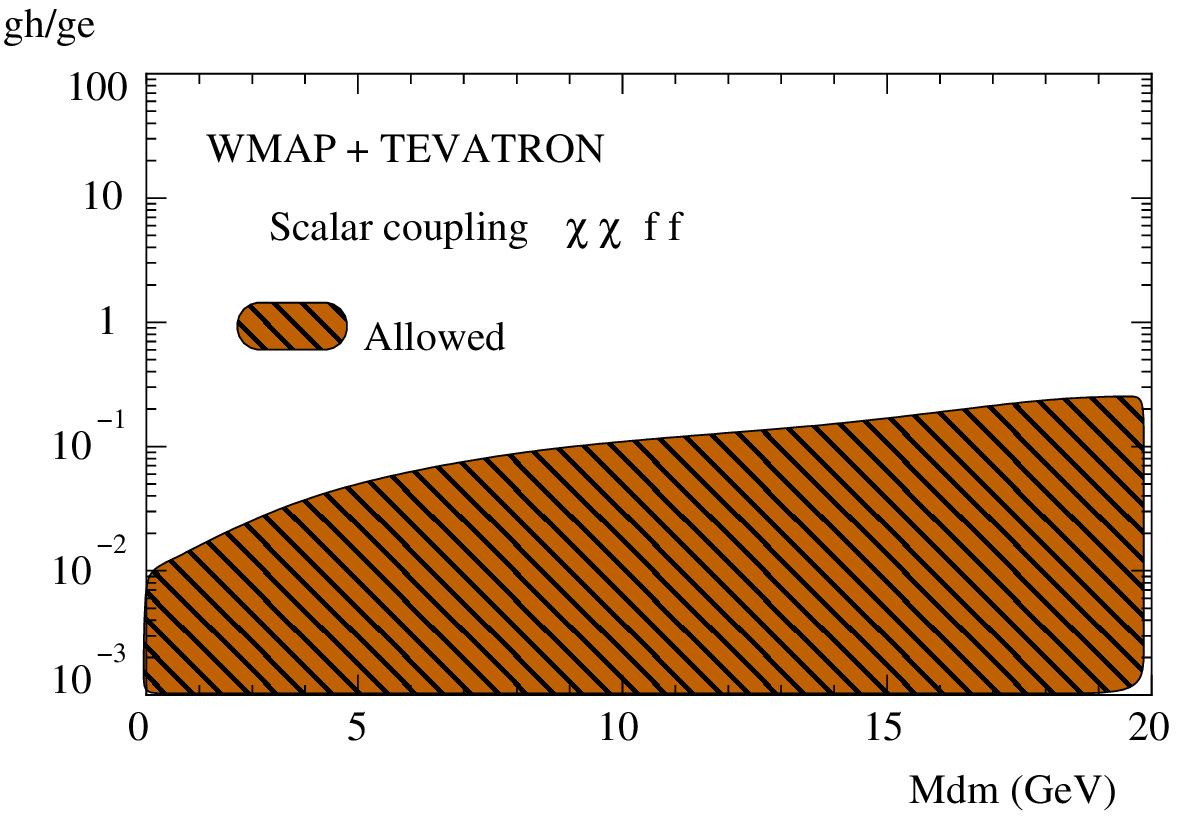}
    
          \caption{{\footnotesize
Hadronic ratio coupling for the annihilation of dark matter as function of the dark matter mass
in the case of electronic couplings for a scalar--like interaction after a scan on $\Lambda_e$
and $\Lambda_h$. After applying the constraint of mono--jet events from LEP (top) and Tevatron (bottom)
 we observe that no point of the parameter space respects both constraints (see the text for details).
}}
\label{Fig:TevatronScalar}
\end{center}
\end{figure}

\noindent
Contrarily to the LEP analysis, the center of mass energy does not limit
the lower bound on $\Lambda_h$ for $m_{\chi}\lesssim 100$ GeV. 
We have only plotted the limit on the up--type coupling, which is the one we used through
the paper  to stay as conservative as possible
(limits of down or charm--type couplings on $\Lambda_h$ are a factor 3 and 10 lower respectively
\cite{Bai:2010hh}).
We can easily understand how the Tevatron constraint imply some strong tensions when combined
with WMAP and LEP analysis. Indeed, to reconciliate LEP constraints with WMAP we needed 
to increase the hadronic contribution (and thus, the coupling to quarks) in the annihilation
process. This then enters in conflict with the limit from the non-observation of mono-jet excess 
at Tevatron. To keep the logic of the work and keep a conservative analysis, we
considered universal leptonic couplings (implying a smaller hadronic contribution
to respect WMAP upper bound).
 The results are shown in Figs.\ref{Fig:TevatronVector} and \ref{Fig:TevatronScalar}.
 Confirming the conclusions of \cite{Bai:2010hh}: for a vector--like coupling
 the Tevatron bound is the more stringent
 for dark matter mass below 5 GeV. Indeed, for such low mass, the hadronic branching ratio
 needed to respect in the meantime WMAP and LEP would produce a clear excess in mono-jet events
 at Tevatron and would have been observed. 
 The Tevatron constraints are even more impressive 
 for a scalar--like couplings, where all the parameter space allowed by the combined WMAP and LEP
 analysis is excluded by Tevatron data (Fig.\ref{Fig:TevatronScalar}).

\subsection{XENON100 constraint}

Recently, the XENON100 collaboration has released several analysis claiming for no detection signal
of dark matter \cite{{Aprile:2011ts}}. Their results are by far the more constraining one 
in the field of direct detection experiments. One can easily understand that XENON100 is
adding new tensions when combined with WMAP, LEP and Tevatron bounds.
Indeed, the hadronic branching fraction required to avoid the overproduction of dark matter 
in the early universe could enter in conflict not only with Tevatron results but also
with XENON100 exclusion limits. Indeed, whereas the s--channel dark matter production
$qq \rightarrow \chi \chi g$ is the process constraining $\Lambda_h$ at Tevatron,
the nuclear recoil gives bound to $\Lambda_h$ (and thus $g_h/g_e$)
 through the t-channel process  $\chi q \rightarrow \chi q$. As we can see
in Fig.\ref{Fig:TevatronVector}, XENON100 restrict even a larger part of the parameter 
space for $m_{\chi}\gtrsim 6$ GeV (which would not have been the case if we took
 into account the previous XENON100 analysis \cite{Aprile:2011hi}). Whereas the scalar--like
 interaction is already excluded without the XENON100 data, dark matter with vector--like coupling
 to the SM still survives in a narrow hadrophilic region  of the parameter 
 space with light dark matter\footnote{which strangely coincides with the CoGENT excess signal 
 \cite{Mambrini:2010dq}},
 and another region definitively hadrophobic for $m_{\chi}\gtrsim 12$ GeV.



\section{Conclusion and prospect}

Recently, several  astrophysical data or would-be signals  has been observed in different dark-matter
oriented experiments.
In each case, one could fit the data at the price of specific nature of the coupling between the Standard Model 
(SM) particles and a light Dark Matter candidate: hadrophilic or leptophilic.
We computed the rate of hadronic coupling needed to respect WMAP combined with 
the LEP and Tevatron constraints
from mono-jet events. We showed that a light fermionic dark matter ($\lesssim 10$ GeV) is mainly excluded whatever is 
its type of interaction, whereas heavier candidates
 ($\gtrsim 20$ GeV) should be largely hadrophobic for vectorial interaction, but excluded for scalar one.
We also studied the special case of scalar dark matter, using the mono--photon events
to constraint the coupling of dark matter to electron with a complete simulation of DELPHI events
and showed that LEP and Tevatron are not able to restrict the couplings.
One of the main consequences is that models with light electrophilic couplings, explaining
INTEGRAL data or constraints from synchrotron radiations are excluded by Tevatron/LEP analysis.
One possibility to escape such strong conclusion would be to suppose that DM has  no electronic coupling. In this case, 
 LEP limits do not apply. Moreover, if at the same time the hadronic coupling is only to the bottom or
 charm quark, Tevatron  XENON100 bounds are not applicable too. However such
 unnatural construction should be excluded by FERMI last analysis
of dwarf galaxies \cite{Garde:2011wr}.


\section*{Acknowledgements}
The authors want to thank particularly T. Schwetz, E. Dudas, A. Falkowski, G. Zaharijas  and F. Bonnet
 for very useful discussions, adding special thanks for JB De Vivie and M. Kado for
 enlightening us with the LEP analysis.  Acknowledgements also to the LPT ``Magic Monday Journal Club" participants for enthusiastic and interesting debates.
  The work was
supported by the french ANR TAPDMS {\bf ANR-09-JCJC-0146} 
and the spanish MICINNÕs Consolider-Ingenio 2010 Programme 
under grant  Multi- Dark {\bf CSD2009-00064}. B. Zald\'\i var acknowledges as well the financial support
of the FPI (MICINN) grant BES-2008-004688, under the contract {\bf FPA-2007-60252}.


\end{document}